\documentclass[prb,aps,twocolumn,amsmath,amssymb,floatfix,superscriptaddress]{revtex4}

\usepackage{graphicx} 
\usepackage{float}
\usepackage{braket}
\usepackage{color}
\definecolor{dred}{rgb}{0.75,0,0}
\usepackage{soul}
\usepackage[colorlinks=true, citecolor=blue, urlcolor=blue ]{hyperref}
\begin{document}

\title{Exact and mean-field analysis of the role of Hubbard interactions on flux driven circular current in a quantum ring}

\author{Rahul Samanta}

\affiliation{Physics and Applied Mathematics Unit, Indian Statistical Institute, 203 Barrackpore Trunk Road, Kolkata-700 108, India}

\author{Santanu K. Maiti}

\email{santanu.maiti@isical.ac.in}

\affiliation{Physics and Applied Mathematics Unit, Indian Statistical Institute, 203 Barrackpore Trunk Road, Kolkata-700 108, India}

\author{Shreekantha Sil}

\affiliation{Department of Physics, Visva-Bharati, Santiniketan, West Bengal-731 235, India}

\begin{abstract}

We investigate circular current in both ordered and disordered Hubbard quantum rings threaded by magnetic flux, employing exact 
diagonalization and the Hartree-Fock mean-field approach within the tight-binding framework. The influence of on-site and extended 
Hubbard interactions, disorder, and electron filling on the persistent current is systematically analyzed. To construct the full 
many-body Hamiltonian, we introduce a linear table formalism, which, to our knowledge, has been rarely used in this context. In ordered 
rings, the current decreases monotonically with increasing on-site repulsion, while the impact of the extended interaction depends 
strongly on the filling factor. At low filling, stronger extended interaction suppresses the current, whereas near half-filling, it 
enhances the current up to a critical ratio, half of the on-site strength, before reducing it. Disorder significantly modifies these 
behaviors, notably enhancing the current at less than quarter-filling with increasing extended interaction. The localization properties 
of eigenstates, examined via the inverse participation ratio, further support the crucial roles of filling and the interplay between 
on-site and extended interactions in governing persistent current.

\end{abstract}

\maketitle

\section{Introduction}
\label{sec:1}

Recent advances in experimental techniques~\cite{e1,e2} enabling the exploration of quantum effects at the mesoscopic scale have 
generated renewed interest in sub-micrometer systems. In this regime, the phase coherence length $L_\phi$ becomes comparable to the 
system size at very low temperatures, leading to discrete energy levels. These conditions are essential for the existence of circular current~\cite{ring1,ring2,ring3} in a ring threaded by magnetic flux. Consequently, small metallic rings at near-zero 
temperature~\cite{zero-temp} have become ideal platforms for investigating mesoscopic quantum phenomena.

The circular current in a quantum ring, induced by an external magnetic flux $\phi$~\cite{ab}, is commonly referred to as the persistent 
current (PC) due to its non-decaying nature even at finite temperature. The existence of PC in disordered metallic rings was first 
proposed by B\"uttiker {\em et al.}~\cite{lb1} and subsequently confirmed by several pioneering experiments~\cite{lb2,lb3}. The 
inclusion of disorder renders the system more realistic but introduces analytical challenges to solve. Despite extensive theoretical 
and experimental efforts to understand this phenomenon, a complete consensus has yet to emerge~\cite{pb1,pb2,pb3}. One persistent 
controversy concerns the magnitude of the persistent current: experiments consistently report currents much larger than those 
predicted theoretically.

In the free-electron model, the maximum current for a one-dimensional perfect ring of length $L$ is $I_0 = 2ev_F/L$, 
where $v_F$ denotes the Fermi velocity~\cite{amp}. The factor of $2$ arises from the spin degeneracy of electrons.
Introducing disorder drastically suppresses this current, whereas experiments reveal values close to $I_0$. 
This discrepancy indicates that the free-electron picture alone is insufficient and that both disorder~\cite{disorder1,disorder2,disorder3} 
and electron-electron interactions must be incorporated for a realistic description. The Hubbard model (HM), which includes nearest-neighbor
hopping (NNH) $t$ and on-site Coulomb repulsion $U$~\cite{ci}, offers a minimal yet powerful framework to study correlation effects. 
Although it is well established~\cite{h1,h2,h3,h4} that the persistent current generally increases with $U$, theoretical predictions 
still fall short of experimental magnitudes.

To reconcile this mismatch, recent studies have extended the HM by including higher-order hopping terms~\cite{hop1,hop2}. However, the 
effect of a more general extended Hubbard model (EHM), which includes both on-site ($U$) and nearest-neighbor ($V$) interactions, on 
the persistent current, particularly in disordered rings, remains insufficiently explored. The addition of the $V$ term, representing
higher-order electron-electron correlations, not only enriches the physical picture but also significantly increases computational 
complexity. The interplay between $U$ and $V$ crucially influences the ground-state energy landscape and, consequently, the magnitude 
of the persistent current (defined as the derivative of the ground-state energy with respect to flux). This behavior is further modulated 
by factors such as disorder strength, system size, and electronic filling. A detailed study of these inter-dependencies is therefore 
essential for bridging the gap between theory and experiment.

The present work is motivated by two primary objectives. First, we aim to provide a clear and accessible framework for constructing 
many-body basis states and Hamiltonian matrix elements, facilitating the implementation of exact diagonalization (ED). Second, we explore 
how filling factor, electron-electron interactions, and disorder collectively determine the circular current in both ordered and 
disordered Hubbard rings.

We employ the tight-binding (TB) formalism~\cite{tb1,tb2,tb3,tb4} to model the system Hamiltonian and analyze it using both exact
diagonalization~\cite{ED1,ED2,ED3,ED4} and Hartree-Fock (HF) mean-field (MF) approaches~\cite{mf1,mf2,mf3,mf4,mf5}. The complete many-body 
Hamiltonian is constructed using a linear (LIN) table formalism~\cite{ED4,LIN1,LIN2,LIN3,LIN4}, {\em enabling efficient analysis even 
for a $10$-site ring up to half-filling}, a case difficult to handle using conventional recursion techniques~\cite{ED5,ED6}. The 
analysis is further extended to larger systems within the mean-field framework. Our results elucidate the distinct roles of the on-site 
($U$) and nearest-neighbor ($V$) interactions and demonstrate that disorder can drastically alter their effects. Notably, we identify 
parameter regimes where the persistent current exhibits significant enhancement with increasing $V$, even for finite $U$. Additionally, 
we compute the inverse participation ratio (IPR)~\cite{ipr1,ipr2} to probe localization properties of eigenstates, which support our 
findings regarding the interplay among $U$, $V$, and filling factor in controlling the persistent current.

The remainder of the paper is organized as follows. Section II presents the theoretical framework, including the construction 
of the system Hamiltonian, the many-body basis via the LIN table formalism, and the computation of persistent current and IPR. 
Section III discusses numerical results for ordered and disordered rings, highlighting the effects of disorder strength and 
Hubbard interactions on the persistent current. Finally, Sec. IV contains the summary of our findings.

\section{Quantum ring, Hamiltonian, and theoretical framework}
\label{sec:2}

\begin{figure}[ht]
\centering
\includegraphics[width=0.85\linewidth]{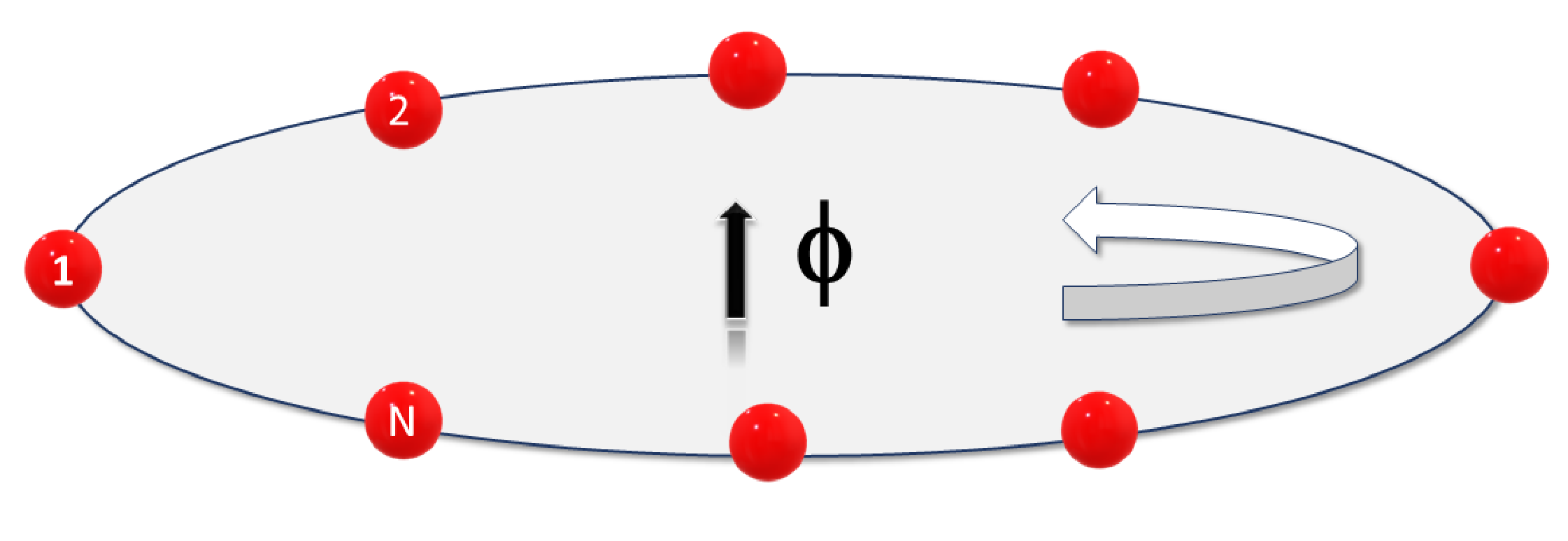}
\caption{(Color online). Schematic diagram of a quantum ring threaded by a magnetic flux $\phi$, where the red spheres represent atomic 
sites in the ring.}
\label{fig:1}
\end{figure}

Figure~\ref{fig:1} shows a schematic representation of the quantum ring, where $N$ denotes the total number of atomic sites. The ring 
is threaded by an Aharonov--Bohm (AB) magnetic flux $\phi$, which induces a circular (persistent) current in the system. 
The flux is measured in units of the elementary flux quantum, $\phi_0=h/e$, where $h$ and $e$ are fundamental constants. In the extended 
Hubbard model, we consider nearest-neighbor hopping, on-site Coulomb repulsion, and nearest-neighbor interaction. The system is described 
within the tight-binding approximation, and the theoretical formulations employed for numerical analysis are detailed below.

\vspace{0.5cm}
\noindent
\textbf{\rule{2.5mm}{2.5mm}} \textbf{Hamiltonian of the AB ring:} Within the TB framework, the Hamiltonian of an $N$-site ring threaded 
by flux $\phi$ is expressed as
\begin{eqnarray}
H &=& \sum_{i,\sigma} \epsilon_{i,\sigma} n_{i,\sigma} + \sum_{i,\sigma} t \left(c_{i,\sigma}^\dagger c_{i+1,\sigma} e^{j\theta} 
+ c_{i+1,\sigma}^\dagger c_{i,\sigma} e^{-j\theta}\right) \nonumber \\
& & + \sum_i U n_{i,\uparrow} n_{i,\downarrow} + \sum_{i,\sigma,\sigma^\prime} V n_{i,\sigma} n_{i+1,\sigma^\prime} \nonumber \\
&= &H_o + H_t + H_U + H_V,
\label{eqn1}
\end{eqnarray}
where, the individual terms of the Hamiltonian $H$ correspond to on-site energy ($H_o$), hopping ($H_t$), on-site Hubbard term ($H_U$), 
and extended Hubbard term ($H_V$). The parameters $\epsilon_i$ and $t$ denote the on-site potential and nearest-neighbor hopping, 
respectively. The operator $c_{i,\sigma}^\dagger$ ($c_{i,\sigma}$) creates (annihilates) an electron with spin 
$\sigma \in \{\uparrow,\downarrow\}$ at site $i$, and $n_{i,\sigma}=c_{i,\sigma}^\dagger c_{i,\sigma}$ is the corresponding number 
operator. The total number operator at site $i$ is thus $n_i = n_{i,\uparrow} + n_{i,\downarrow}$. In the presence of magnetic flux 
$\phi$, a phase factor $\theta = 2\pi\phi/(N\phi_0)$ appears in the hopping term. 

The Hamiltonian is solved both exactly, using the exact diagonalization (ED) method, and using the Hartree--Fock (HF) mean-field 
approach, as discussed below.

\vspace{0.5cm}
\noindent
\textbf{\rule{2.5mm}{2.5mm}} \textbf{Exact diagonalization of full many-body Hamiltonian:} A key step in this prescription is the 
construction of the full many-body Hamiltonian using an appropriately chosen basis. Each lattice site can have four possible options: 
$\{0, \uparrow, \downarrow, \uparrow\downarrow\}$, resulting in a total Hilbert space dimension of $4^N$, for an $N$-site ring system. 
To reduce this size, we fix the total number of electrons and separately constrain the numbers of up- and down-spin electrons, 
{$N_\uparrow$, $N_\downarrow$}, reducing the dimension to ${}^{N}C_{N_\uparrow} \times {}^{N}C_{N_\downarrow}$.

To efficiently construct and index all basis states, we employ the \textit{LIN table formalism}~\cite{ED4,LIN1,LIN2,LIN3,LIN4}, 
which uses binary (bit-string) representations for occupation configurations. Each basis state in a spin subspace $\sigma$ is 
represented by a bit-string $b_\sigma$ of length $N$, containing $N_\sigma$ ones (occupied sites) and $N-N_\sigma$ zeros 
(unoccupied sites). The set of all possible combinations forms the complete spin sub-space
\[B_{\sigma} = \{b_\sigma^{(0)}, b_\sigma^{(1)}, \ldots, b_\sigma^{({}^{N}C_{N_\sigma}-1)}\}.\]
Each full many-body basis state is then represented as
\begin{equation}
\text{Basis} = \{(b_\uparrow^u, b_\downarrow^d) \mid b_\uparrow^u \in B_\uparrow,\, b_\downarrow^d \in B_\downarrow\},
\end{equation} 
where $u$ and $d$ denote the up-spin and down-spin sector indices, respectively, and we assign a linear index to each pair $(u,d)$ 
using the rule $k = {}^{N}C_{N_\uparrow} \times u + d$.

As an illustrative example, here we consider a $3$-site ring with two up-spin and one down-spin electrons. 
Tables~\ref{tab:Lin table} and \ref{tab: LinTable for index} demonstrate how the spin sub-spaces and linear indices are constructed.
\begin{table}[ht]
\centering
\caption{Bit-string representation and corresponding indices for the up- and down-spin subs-paces.}
\vskip 0.2cm
\begin{tabular}{|c|c|c|c|c|c|}
		\hline
		\multicolumn{3}{|c|}{Up-spin} & \multicolumn{3}{c|}{Down-spin} \\ \hline
		$u$ & Bit pattern & Bit value & $d$ & Bit pattern & Bit value \\ \hline
		0 & 011 & 3 & 0 & 001 & 1 \\
		1 & 101 & 5 & 1 & 010 & 2 \\
		2 & 110 & 6 & 2 & 100 & 4 \\ \hline
\end{tabular}
\label{tab:Lin table}
\end{table}

The diagonal elements of the Hamiltonian matrix are obtained from the on-site energy, on-site interaction, and extended Hubbard interaction
terms. For any state $|k\rangle_s = |b_\uparrow^u, b_\downarrow^d\rangle_b$, these contributions are given by
\begin{align}
\langle H_o \rangle &= \sum_i \epsilon_i \big[(b_\uparrow^u)_i + (b_\downarrow^d)_i\big], \\
\langle H_U \rangle &= \sum_i U \big[(b_\uparrow^u)_i \cdot (b_\downarrow^d)_i\big], \\
\langle H_V \rangle &= \sum_i V \big[(b_\uparrow^u)_i + (b_\downarrow^d)_i\big]\big[(b_\uparrow^u)_{i+1} + (b_\downarrow^d)_{i+1}\big].
\end{align}

Off-diagonal elements arise from the hopping term, where single-bit shifts within a spin sub-space correspond to electron motion. 
\begin{table}[ht]
\centering
\caption{Linear indexing of the many-body basis states.}
\vskip 0.2cm
\begin{tabular}{|c|c|} \hline
Combination $(u,d)$ & State index $k$ \\ \hline
(0,0) & 0 \\ (0,1) & 1 \\ (0,2) & 2 \\ (1,0) & 3 \\ (1,1) & 4 \\ (1,2) & 5 \\ (2,0) & 6 \\ (2,1) & 7 \\ (2,2) & 8 \\ \hline
\end{tabular}
\label{tab: LinTable for index}
\end{table}
The effect of flux is included through factors $e^{\pm j\phi}$ for forward and backward hoppings, respectively, and the 
fermionic signs are handled using $\text{sign} = (-1)^{N_\sigma - 1}$ for the periodic boundary condition.

The resulting Hamiltonian matrix is diagonalized using the Lanczos algorithm, implemented through the ARPACK linear algebra 
library~\cite{Lanczos1, Lanczos2}, which efficiently computes the lowest eigenvalues corresponding to the ground-state (GS) energy.

\vspace{0.5cm}
\noindent
\textbf{\rule{2.5mm}{2.5mm}} \textbf{Mean--field approach:}
Although exact diagonalization of the full many-body Hamiltonian is the most reliable approach for obtaining the energy eigenvalues, 
and hence the ground state energy and the flux-induced circular current, its computational cost grows exponentially with system size, 
making it impractical for larger systems. To analyze larger rings, we therefore employ the mean-field (MF) method~\cite{mf1,mf2,mf3,mf4}, 
which decouples the interaction terms in Eq.~\ref{eqn1} into effective one-body contributions. The resulting Hamiltonian can be written as
\begin{equation}
\begin{aligned}
H &= H_\uparrow + H_\downarrow - \sum_i \Big[ U \langle n_{i,\uparrow}\rangle \langle n_{i,\downarrow}\rangle 
+ V \sum_{\sigma,\sigma'} \langle n_{i,\sigma}\rangle \langle n_{i+1,\sigma'}\rangle \Big],
\end{aligned}
\label{eqn2}
\end{equation}
where $H_{\uparrow}$ and $H_{\downarrow}$ correspond to the up and down spin Hamiltonians, respectively, and $\langle n_{i,\sigma}\rangle$
denotes the occupation probability. The effective site energies for the two spin cases become
\begin{equation}
\epsilon_{i,\uparrow(\downarrow)}^{\text{eff}} = \epsilon_i + U \langle n_{i,\downarrow(\uparrow)} \rangle + V \sum_{j,\sigma} \langle n_{j,\sigma} \rangle,
\label{eqn3}
\end{equation}
and the explicit forms of $H_{\uparrow}$ and $H_{\downarrow}$ are
\begin{eqnarray}
H_{\uparrow(\downarrow)} & = & \sum_i \epsilon_{i,\uparrow(\downarrow)}^{\text{eff}} c_{i,\uparrow(\downarrow)}^\dagger 
c_{i,\uparrow(\downarrow)} \nonumber \\
 & + &  t \left(c_{i,\uparrow(\downarrow)}^\dagger c_{i+1,\uparrow(\downarrow)} e^{j\theta} + h.c.\right).
\label{eqn4}
\end{eqnarray}

\vspace{0.5cm}
\noindent
\textbf{Self-consistency:}
Starting from initial guess values for $\langle n_{i,\uparrow}\rangle$ and $\langle n_{i,\downarrow}\rangle$, 
labeled as $\langle n_{i,\uparrow}^{\text{old}}\rangle$ and $\langle n_{i,\downarrow}^{\text{old}}\rangle$, we 
construct $H_\uparrow$ and $H_\downarrow$. Upon diagonalization, we obtain new occupation probabilities
$\langle n_{i,\uparrow(\downarrow)}^{\text{new}}\rangle = \sum_{k=1}^{n_{\uparrow(\downarrow)}} |\psi_i^{(k)}|^2$,
which are iteratively updated until a converged solution is achieved. We define the convergence by a factor
$$\Delta S_i=(\langle n_{i,\uparrow}^{\text{new}}\rangle-\langle n_{i,\uparrow}^{\text{old}}\rangle)^2+(\langle n_{i,\downarrow}^{\text{new}}\rangle-\langle n_{i,\downarrow}^{\text{old}}\rangle)^2$$
and stop the iteration when $\Delta S_i<10^{-5}.$

\vspace{0.5cm}
\noindent
\textbf{Ground-state energy and persistent current:} At zero temperature ($T=0\,$K), the ground-state energy for a given filling 
is obtained as
\begin{equation}
\begin{aligned}
E_g &= \sum_p \left(E_{p,\uparrow} + E_{p,\downarrow}\right) - \sum_i \Big[ U \langle n_{i,\uparrow}\rangle \langle 
n_{i,\downarrow}\rangle \\
&\quad + V \sum_{\sigma,\sigma'} \langle n_{i,\sigma}\rangle \langle n_{i+1,\sigma'}\rangle \Big],
\end{aligned}
\label{eqn5}
\end{equation}
where $E_{p,\uparrow(\downarrow)}$ are the single-particle eigenvalues of $H_{\uparrow(\downarrow)}$. 
Here, it is relevant to point out that the occupation probabilities, obtained from the eigenstates, depend on the TB 
parameters of the Hamiltonian, and thus any change in these parameters leads to a modification of $\langle n_{i,\sigma}\rangle$, resulting 
in a variation of $E_g$.

The persistent current is calculated from the flux derivative of $E_g$~\cite{pc1,pc2,pc3} as
\begin{equation}
I(\phi) = -\frac{\partial E_g(\phi)}{\partial \phi}.
\label{eqn6}
\end{equation}

\vspace{0.5cm}
\noindent
\textbf{\rule{2.5mm}{2.5mm}} \textbf{Inverse participation ratio:}
To analyze localization behavior, we compute the inverse participation ratio (IPR)~\cite{ipr1,ipr2} for each normalized eigenstate 
$\ket{\phi_k} = \sum_i \psi_i^k \ket{i}$ following the prescription
\begin{equation}
\text{IPR}_k = \sum_{i=1}^{N} |\psi_i^k|^4.
\label{eqn7}
\end{equation}
An extended state corresponds to $\text{IPR}_k \to 0$, while a localized state gives $\text{IPR}_k \to 1$. Here it is important to note 
that, these limiting values ($0$ and $1$) are obtained only in the asymptotic limit ($n\rightarrow \infty$). However, for finite systems,
the conducting behavior can still be inferred from comparatively low or high values of $\text{IPR}_k$.

\section{Numerical results and discussion}
\label{sec:3}

\begin{figure*}[ht]
{\centering \resizebox*{17cm}{15.5cm}{\includegraphics{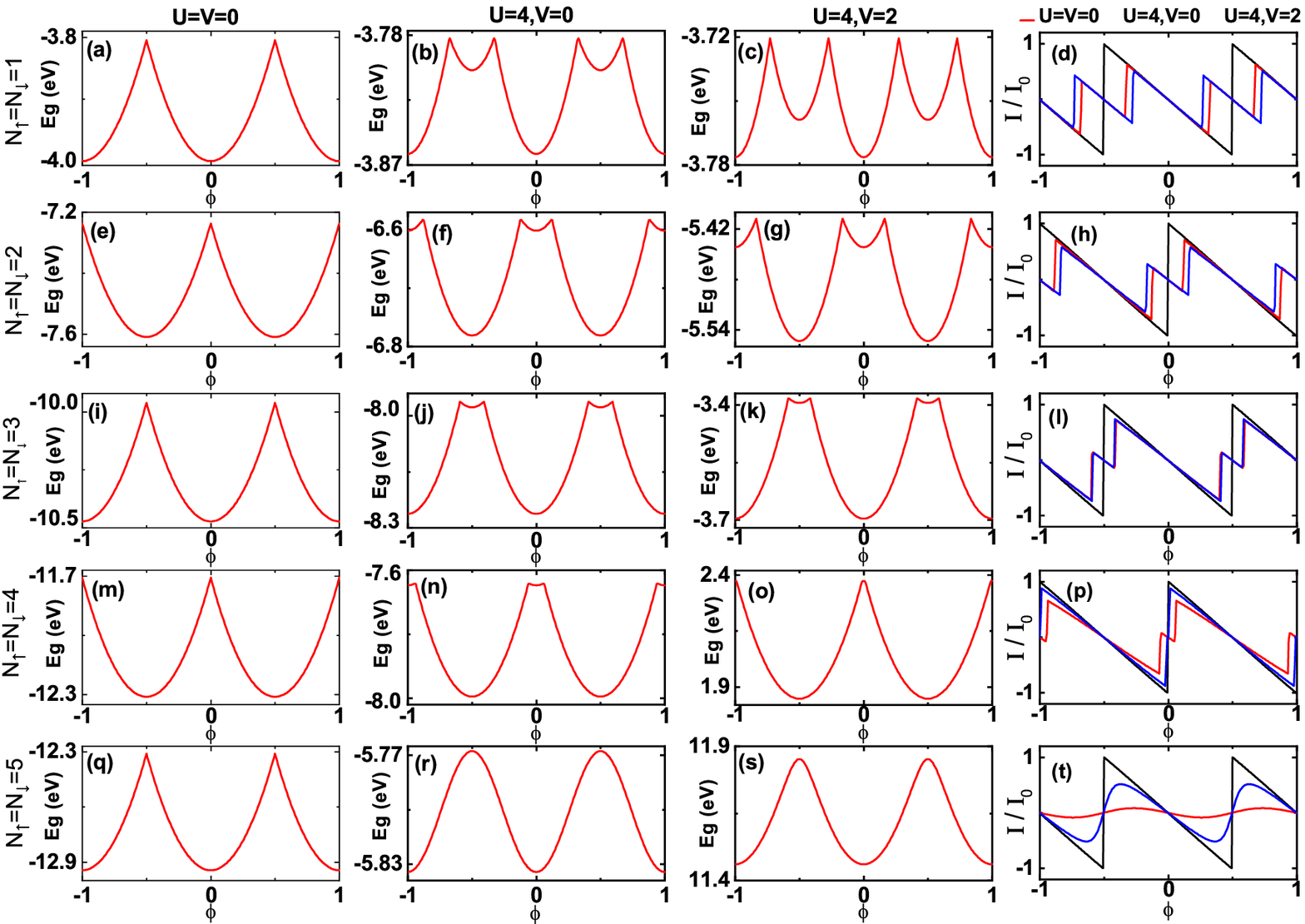}}\par}
\caption{(Color online). Variations of the ground-state energy and the corresponding persistent current as a function of 
AB flux $\phi$ (measured in units of $\phi_0$) for an ordered ring with $N=10$. The first three columns present the ground-state energy 
for the cases $U=V=0$; $U=4,\,V=0$; and $U=4,\,V=2$, respectively, while the associated persistent currents are displayed in the last column. Beginning with $N_\uparrow = N_\downarrow = 1$ in the top row, the numbers of both $N_\uparrow$ and $N_\downarrow$ electrons are increased by one in each subsequent row until the half-filled configuration is reached in the bottom row.}
\label{N=10, order ring}
\end{figure*}

To investigate the interplay between on-site and extended Hubbard interactions on the flux-driven circular current, in this section, we present and critically analyze numerical results under various input conditions. The results are organized into two parts. The first part focuses on those obtained from exact diagonalization of the full many-body Hamiltonian, which always yields a more accurate descriptions. In this approach, within our computational facilities, we are able to study rings with a maximum size of $10$ sites (up to the half-filled regime), considering different combinations of up- and down-spin electrons. The second part presents the mean-field results, used to explore larger ring systems.
The accuracy of the mean-field results is checked for the sake of completeness of our analysis.

All characteristics of the persistent current are investigated at absolute zero temperature. Throughout the calculations, we set $t=1\,$eV, and all energies are expressed in units of $t$. The currents are measured in units of $I_0$. The remaining physical parameters are specified at the appropriate places in the text.

\subsection{Exact diagonalization analysis}

\subsubsection{Ordered quantum ring}

The ordered mesoscopic ring corresponds to a quantum ring free from any impurities. Thus, without loss of generality, we set the on-site potential at each site to zero ($\epsilon_i=0$). As mentioned, all the nearest-neighbor hopping amplitudes are set at $1\,$eV. In this sub-section, we explore the influence of on-site and extended Hubbard interactions on the ground-state energy $E_g$ and the persistent current in ordered rings.

Figure~\ref{N=10, order ring} presents the variation of the ground-state energy and PC for a $10$-site ordered ring under different electron fillings and interaction strengths. Starting from $N_{\uparrow}=N_{\downarrow}=1$ in the first row, the number of electrons in each spin sector is increased by one in each subsequent row until the half-filled band limit is reached. The first three columns of the figure show the ground-state energy $E_g(\phi)$ for $(U,V)=(0,0)$, $(U,V)=(4,0)$, and $(U,V)=(4,2)$, respectively. The corresponding PC variations over the two-flux-quantum range are displayed in the fourth column. Below, we analyze the results one by one.

\vspace{0.2cm}
\noindent
\textbf{Characteristics of ground-state energy:} A careful examination of the $E_g$-flux curves reveals several important features those are as follows. (i) Sharp peaks appear at $\phi = 0$ and $\phi = m\phi_0/2$ ($m$ is an integer), accompanied by abrupt changes in the slope of the energy curves. These slope variations signal qualitative changes in the nature of the ground state. (ii) The inclusion of the on-site Hubbard interaction $U$ leads to the emergence of additional local minima in $E_g(\phi)$. Our numerical results indicate that the width of these minima increases with increasing $U$. (iii) When the extended Hubbard interaction $V$ is added on top of $U$, both the 
width and depth of the local minima increase for low fillings (up to the third row). However, as the system approaches half-filling, the inclusion of $V$ reduces the width of the dip region. (iv) For an ordered ring in the non-interacting case, $E_g(\phi)$ shows sharp slope changes at isolated flux points. In contrast, for a disordered ring, these slope variations become smooth. For the non-interacting systems, these features have already been studied. Interestingly, in our case, even for the ordered ring with $N_{\uparrow}=N_{\downarrow}=5$, the inclusion of both $U$ and $V$ leads to a smoothening of the $E_g$ slope, which serves as a clear signature of the interplay between these two interactions. (v) The interaction parameters $U$ and $V$, which modify the flux-dependent ground-state energy $E_g$, 
are repulsive, and as a consequence, the magnitude of $E_g(\phi)$ increases with increasing $U$ and $V$.

Now, we try to explain these characteristics features. In the absence of interactions, the ground-state energy is simply the sum of occupied single-particle energy levels. The crossings of these levels at $\phi=0$ and $\phi=m\phi_0/2$ give rise to the cusps observed in the first column of Fig.~\ref{N=10, order ring}. The reordering of levels due to these degeneracies results in abrupt slope changes.

The on-site Hubbard term contributes only for doubly occupied sites. Consequently, the energies of many-body states without double occupancy remain independent of $U$, whereas those with double occupancy shift with increasing $U$. As $U$ increases, the relative shifting of these many-body levels generates additional crossings, which manifest as new local minima in $E_g(\phi)$. A larger $U$ pushes doubly occupied configurations to higher energy, shifting the flux positions of these crossings and thereby broadening the associated minima.

When the extended e-e interaction $V$ is included along with $U$, the system energetically penalizes the simultaneous occupancy of nearest-neighbor sites. At low fillings, the presence of many empty sites allows electrons to avoid both double occupancy and nearest-neighbor repulsion, which stabilizes extended minima in $E_g(\phi)$. As the filling approaches half, the electrons can no longer fully avoid nearest-neighbor occupation, leading to a competition between the effects of $U$ and $V$. Consequently, the tendency of the system to settle into a local minimum is reduced, leading to a narrowing of its width.

Inspecting the plots in the first three columns of each row, we clearly observe that the magnitude of $E_g$ increases as the interaction strengths are introduced. Specifically, for any value of the AB flux $\phi$, the magnitude of $E_g$ in the second column corresponding to $U=4$, $V=0$ is larger than that in the non-interacting case ($U=V=0$) shown in the first column. This magnitude increases further when the nearest-neighbor interaction $V$ is included along with $U$, as illustrated in the third column for the case $U=4$, $V=2$. 
The observed enhancement in the magnitude of $E_g$, at any flux, with increasing $U$ and $V$ is consistent with the system Hamiltonian 
given in Eq.~\ref{eqn1}, where both interaction terms appear with positive coefficients. The on-site Hubbard interaction
contributes through double occupancy, while the extended Hubbard interaction contributes through nearest-neighbor density correlations. 
Since these contributions originate from the flux-dependent many-body Hamiltonian, the resulting enhancement in $E_g$ becomes flux dependent,
which is also evident from the presented $E_g$-$\phi$ plots.

\begin{figure*}[ht]
{\centering \resizebox*{16cm}{11cm}{\includegraphics{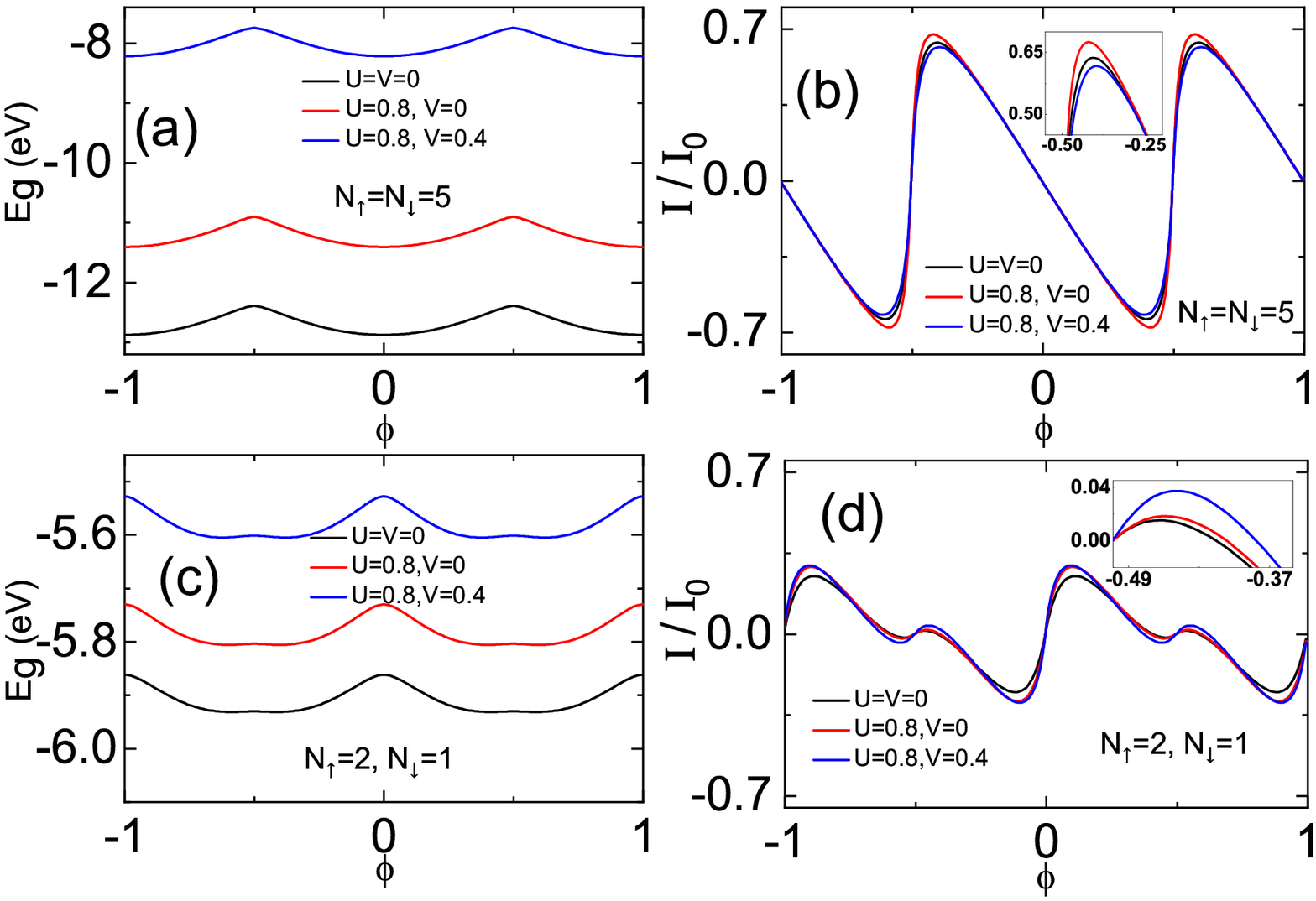}}\par}
\caption{(Color online). Variations of the ground-state energy and persistent current with magnetic flux $\phi$ (measured 
in units of $\phi_0$) for a $10$-site disordered ring 
(random disorder) with disorder strength $W=1$. The left and right columns display the flux dependence of $E_g$ and PC, respectively. 
The top row corresponds to the half-filled configuration ($N_{\uparrow} = N_{\downarrow} = 5$), while the bottom row represents the less-than-quarter-filled case ($N_{\uparrow} = 2$, $N_{\downarrow} = 1$). In sub-figures (a) and (c), the black, blue, 
and red curves depict the variation of $E_g$ for $U=V=0$; $U=0.8,\,V=0$; and $U=0.8,\,V=0.4$, respectively. Sub-figures (b) and (d) 
show the corresponding PC variations using the same color scheme in their respective filling regime. Insets are used to highlight the 
current variations more clearly.}
\label{N=10, random disorder ring}
\end{figure*}

\vspace{0.2cm}
\noindent
\textbf{Current-flux characteristics:} For an ordered non-interacting ring, the magnitude of the PC increases as the number of electrons increases, reaching its maximum in the half-filled band. This trend is evident in the fourth column of Fig.~\ref{N=10, order ring}, where although the currents are presented in normalized unit, the peak value of corresponding current $I$ 
increases from $0.77\,\mu\text{A}$ for $N_{\uparrow}=N_{\downarrow}=1$ to $2.51\,\mu\text{A}$ for $N_{\uparrow}=N_{\downarrow}=5$. 
This behavior arises because, in the low-filling regime, the abundance of empty sites allows added 
electrons to delocalize easily, thereby enhancing their contribution to the current. As the system approaches half-filling, the number of accessible hopping sites decreases, reducing the rate of increase in the current. At half-filling, the current reaches its maximum, and the system attains its highest conducting state. Beyond this point, the current gradually decreases and eventually drops to zero as the ring approaches full filling.

Introducing the Hubbard interaction $U$ restricts electron motion, thereby reducing the current for all fillings. Moreover, the rate of suppression increases as the system approaches half-filling. For $U=4$, the reduction in the peak value of the current $I$
from the non-interacting value increases progressively from the first to the last row, with differences of $0.312$, $0.557$, $0.700$, $1.055$, and $2.314\,\mu\text{A}$, respectively. This clearly demonstrates that the effect of $U$ becomes more severe near half-filling. This behavior may be understood as follows. 
At low fillings, $U$ primarily suppresses electron hopping. Near half-filling, however, $U$ additionally promotes
spin-density-wave (SDW) correlations, characterized by an alternating up-down spin arrangement, which practically localizes electrons 
to specific lattice sites and substantially reduces the current, leading to an insulating phase in the system.
This localization phenomenon in the half-filled limit can also be well understood in terms of the band structure. The spin-density-wave 
(SDW) ordering implies a binary lattice-type modulation, which leads to the formation of two energy sub-bands separated by a gap. This 
gap is associated with $U$ and increases with increasing $U$. At zero temperature, in the half-filled limit, the highest occupied energy 
level lies at the top of the lower sub-band, while the upper sub-band remains completely unoccupied. Since the lower band is fully filled, 
the net circular current vanishes, indicating the insulating phase.

The effect of adding the extended interaction $V$ on top of $U$ is more nuanced. As $V$ increases from $0$ to $2$, the current initially decreases for $N_{\uparrow}=N_{\downarrow}=1$ to $3$. For $N_{\uparrow}=N_{\downarrow}=4$, the current increases slightly, and for $N_{\uparrow}=N_{\downarrow}=5$ it increases substantially. Our detailed analysis reveals that, at half-filling, a significant drop in current reappears when $V > U/2$.
For low fillings, $V$ imposes an additional constraint on electron motion, leading to further suppression of the current (as seen in Figs.~\ref{N=10, order ring}(d) and (h)). Near half-filling, all sites become occupied, and the system experiences competing effects: $U$ favors alternating-spin occupation on neighboring sites, while $V$ penalizes such nearest-neighbor occupancy. Increasing $V$ beyond a certain threshold weakens the pinning induced by $U$, promoting partial delocalization and resulting in an enhanced current. Our analysis shows that the current reaches a maximum near $V = U/2$. However, when $V$ exceeds $U/2$, the system again tends toward localization. In this 
regime, it favors a charge-density-wave (CDW) configuration, where opposite-spin electrons are pinned to the same site, leading to a 
reduction in the current.

From the behavior of the above studied currents under different values of $U$ and $V$, we identify distinct phases: a highly conducting 
phase, an SDW insulating phase, and a CDW phase. A more detailed description of these phases, along with additional ones, can be found 
in Refs.~\cite{phase_diagram1, phase_diagram2}, in the presence of both on-site and extended Hubbard interactions, as considered in our 
present work.

\subsubsection{Disordered quantum ring}

In general, disorder can be of two types: random and correlated. In random disorder, there is no spatial correlation between the values, whereas correlated disorder exhibits some degree of long-range correlation. In a quantum ring, such disorder can appear in several ways, for example, in the on-site energies, in the hopping amplitudes, or in both. In our work, we consider disorder only in the site energies for simplicity, and we critically analyze the results for both types of disorder to examine whether any special features emerge depending on whether the disorder is correlated or not.

\vspace{0.5cm}
\noindent
\textbf{\textit{2.1 Random disorder}}
\vspace{0.2cm}

In this sub-section, we introduce disorder in the on-site potentials, where each site energy is chosen randomly~\cite{al} from the interval 
[$-0.5W, 0.5W$] with $W$ denoting the disorder strength. The disordered potentials are taken to be identical for both spin species. 
In a randomly disordered ring, the chosen realization of on-site potentials plays a crucial role, together with the interplay of on-site 
and extended Hubbard interactions, in determining the characteristics of the ground-state energy and the persistent current. 
As the ring is randomly disordered, we compute all results by averaging over a large number ($90$) of distinct 
configurations. Increasing the number of configurations further does not lead to any appreciable change ($<10^{-6}$) in the results.

\vspace{0.2cm}
\noindent
\noindent\textbf{Energy-flux characteristics:} Figure~\ref{N=10, random disorder ring}(a) shows the variation of the ground-state energy 
$E_g$ as a function of the AB flux $\phi$ for a mesoscopic ring of size $N=10$ at half-filling ($N_{\uparrow}=N_{\downarrow}=5$), while
Fig.~\ref{N=10, random disorder ring}(c) presents the same for a less-than-quarter-filled configuration ($N_{\uparrow}=2, N_{\downarrow}=1$). 
Here, the disorder strength is fixed at $W=1$. In both Figs.~\ref{N=10, random disorder ring}(a) and 
~\ref{N=10, random disorder ring}(c), the black curve corresponds to the non-interacting limit $U=V=0$. The red curve denotes the 
on-site interacting case ($U=0.8, V=0$), while the blue curve represents the situation where both the on-site and
nearest-neighbor interactions are present ($U=0.8, V=0.4$).

Two general features clearly emerge: (i) even in the absence of interactions, the introduction of disorder produces a continuous, smooth
oscillation of $E_g$ with $\phi$, in sharp contrast to the behavior of an ordered ring. This smooth variation persists upon including
interactions. (ii) For any filling and any value of $\phi$, the magnitude of $E_g$ increases once interactions are switched on, and it 
continues to rise with increasing interaction strength. These observations can be understood from the symmetry consideration. In an 
ordered ring, translational symmetry ensures that the states can be labeled by total momentum. At specific flux values, these 
symmetry-protected states become exactly degenerate. As a result, the ground state switches abruptly between these states, producing 
sharp level crossings in the $E_g$-$\phi$ characteristics. Introducing on-site disorder breaks translational symmetry, so momentum is 
no longer a good quantum number. The many-body states mix, lifting degeneracies and replacing abrupt crossings with smooth reshaping 
of the ground state as the flux varies. Consequently, the cusp-like features in the ordered case are replaced by rounded, smooth 
variations of $E_g$ in the disordered system. The computational results also confirm that the ground-state energy increases with both 
on-site ($U$) and nearest-neighbor ($V$) Hubbard interaction strengths. This behavior is fully consistent with the system Hamiltonian 
(Eq.~\ref{eqn1}), and can be understood following the discussion presented earlier for the ordered ring. 

The rate at which the slope of $E_g$ changes with flux depends sensitively on the interaction strengths $U$ and $V$ as well as 
on the filling factor. These dependencies manifest directly in the behavior of the persistent current, whose analysis is presented 
in the next sub-section.

\vspace{0.2cm}
\noindent
\textbf{Current-flux characteristics:} The persistent current in a mesoscopic ring exhibits several rich and subtle features, and 
its behavior in the presence of disorder is particularly intriguing. In Figs.~\ref{N=10, random disorder ring}(b) and (d), we present the variation of PC as a function of the AB flux $\phi$ for two different filling factors. Figure~\ref{N=10, random disorder ring}(b) shows the current-flux characteristics for the half-filled case at disorder strength $W=1$. The black curve corresponds to the non-interacting limit ($U=V=0$), the red curve reflects the effect of on-site interaction ($U=0.8, V=0$), and the blue curve represents the case where both $U$ and $V$ are finite ($U=0.8, V=0.4$).

A few generic features emerge from these results. First, the magnitude of the persistent current gets reduced compared to the ordered case. This reduction is a direct consequence of disorder-induced localization, which tends to trap electrons at specific sites and thereby suppress their mobility. The inclusion of on-site interaction $U$, however, partially counteracts this localization effect. Since the on-site Hubbard repulsion disfavors double occupancy, it effectively enhances the delocalization tendency and increases the overall kinetic energy of the system, leading to an enhancement of the PC. When nearest-neighbor interaction $V$ is added, the current amplitude decreases again. To minimize nearest-neighbor repulsion in a half-filled ring, electrons attempt to occupy the same site in the absence of sufficiently free neighboring sites, which reduces their hopping probability and consequently suppresses the PC.

Figure~\ref{N=10, random disorder ring}(d) illustrates the corresponding PC variation for the less-than-quarter-filled case with $N_\uparrow = 2$ and $N_\downarrow = 1$. Here, the black, red, and blue curves represent the cases ($U=V=0$), ($U=0.8, V=0$), and ($U=0.8, V=0.4$), respectively. As in the half-filled case, the circular current increases with increasing on-site interaction $U$, owing to the enhancement in the average kinetic energy arising from reduced double occupancy.

In contrast to the half-filled situation, the role of the nearest-neighbor interaction $V$ is markedly different in this dilute regime.
With plenty of empty neighboring sites available, the extended Hubbard repulsion can push electrons away from each other without confining them to the same site. This process increases their mobility and therefore enhances the current. Consequently, in the less-than-quarter-filled band, the PC increases with the strength of the nearest-neighbor interaction $V$, in agreement with the numerical results.

\vspace{0.5cm}
\noindent
\textbf{\textit{2.2 Correlated disorder}}
\vspace{0.2cm}

We know that random disorder is uncorrelated in nature, that simply implies that there is no relation between 
\begin{figure}[ht]
{\centering \resizebox*{8cm}{10cm}{\includegraphics{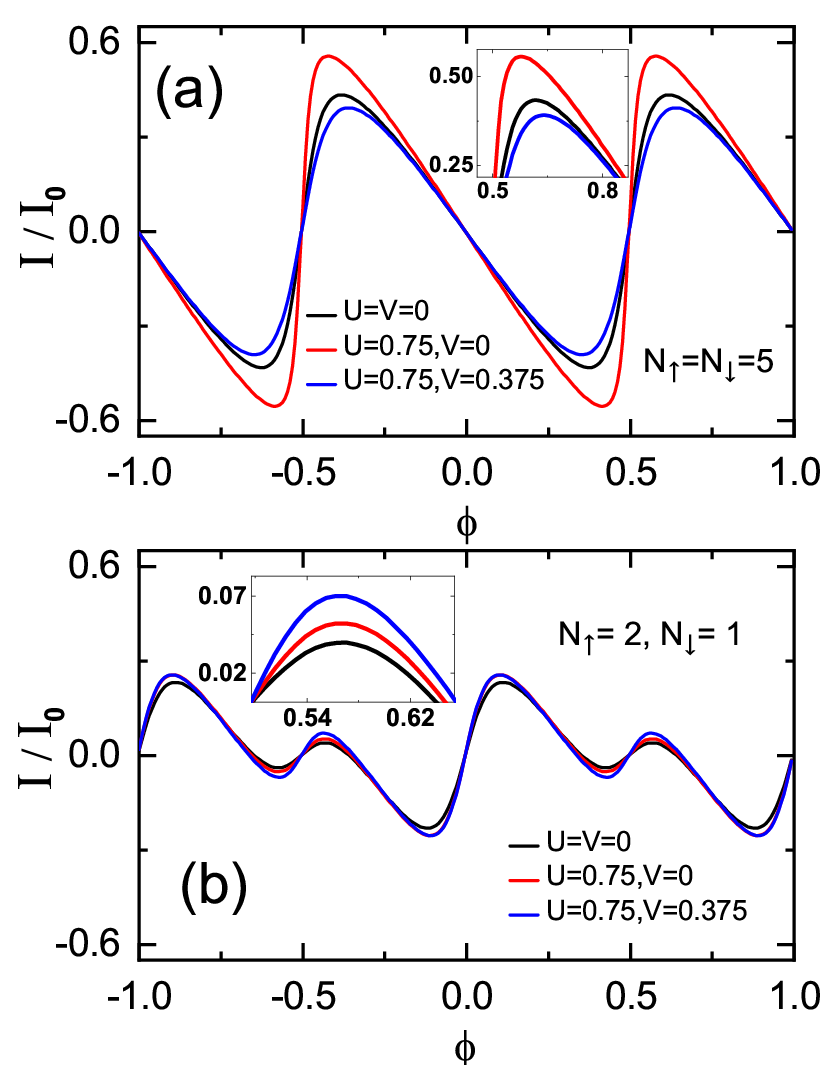}}\par}	
\caption{(Color online). Dependence of the persistent current on the magnetic flux $\phi$ (measured in units of $\phi_0$)
for a correlated disordered ring ($W=1$) of size $N=10$. The black, red, and blue curves represent the PC for 
$U=V=0$; $U=0.75$ with $V=0$; and $U=0.75,\,V=0.375$, respectively, under the half-filled band condition in sub-figure (a) and 
under a less than quarter-filled condition $(N_{\uparrow}=2$, $N_{\downarrow}=1)$ in sub-figure (b). Insets are used to highlight 
the current variations more clearly.}
\label{N=10, correlated disorder ring}
\end{figure}
potentials at any two sites. 
There is another type of disorder in which though the site potentials are deterministic but aperiodic in nature, this type of disorder is 
called as quasi-periodic disorder. Aubry-Andr\'e-Harper (AAH)~\cite{aah1,aah2} model is a well-known example of such kind of disorder, that 
is deterministic, non-periodic yet exhibits a long range ordering. In this model the on-site potential is defined as 
$\epsilon_i = W \cos(2\pi \beta i )$, where we set $\beta = (1+\sqrt{5})/2.$

It is clearly seen from Fig.~\ref{N=10, correlated disorder ring}(a) that in a half-filled ($N_{\uparrow}=N_{\downarrow}=5$) correlated disordered ring of size $N=10$, the current gets enhanced when on-site interaction strength $U$ becomes $0.75$ with respect to non-interacting scenario. Then, the current reduces again with further introduction of $V$ of strength $0.375$ in presence of $U$. This is what reflected by the curve in the figure, where the black curve stands for $U=V=0$, red curve reflects $U=0.75, V=0$ and blue curve shows $U=0.75, V=0.375$ case. In Fig.~\ref{N=10, correlated disorder ring}(b) the black, red and blue lines represent the variation of PC  for the same set of parameters in the less-than-quarter-filled regime ($N_{\uparrow}=2, N_{\downarrow}=1$).
This indicates that here also with increasing on-site interaction strength the current enhances however with further application of nearest-neighbor repulsion strength, the current increases again.

This observation is fully consistent with our previous discussion, which clearly indicates that the magnitude of the 
persistent current depends on several factors, including the filling factor and the strengths of both on-site and extended Hubbard interactions. Regardless of the filling, the on-site interaction $U$ always enhances the current in a disordered ring. In contrast, the role of the nearest-neighbor interaction $V$ varies with the filling: for low filling (less than quarter-filled case), the current increases with increasing $V$, whereas for higher filling (greater than quarter-filled band case), it decreases as $V$ becomes stronger. The results presented in Fig.~\ref{N=10, correlated disorder ring} further support our analysis, demonstrating that our findings remain robust in the presence of both correlated and uncorrelated disorder.

\subsection{Mean-field analysis}

In this part, we present and analyze the results obtained by solving the Hamiltonian within the HF mean-field framework. Our primary 
objective is to examine whether the characteristic features observed from the exact diagonalization of the full many-body Hamiltonian 
remain valid when the system size is increased and treated under the MF approximation. The ground-state energy is calculated using
Eq.~\ref{eqn5}, and 
\begin{figure}[ht]
{\centering \resizebox*{8cm}{9cm}{\includegraphics{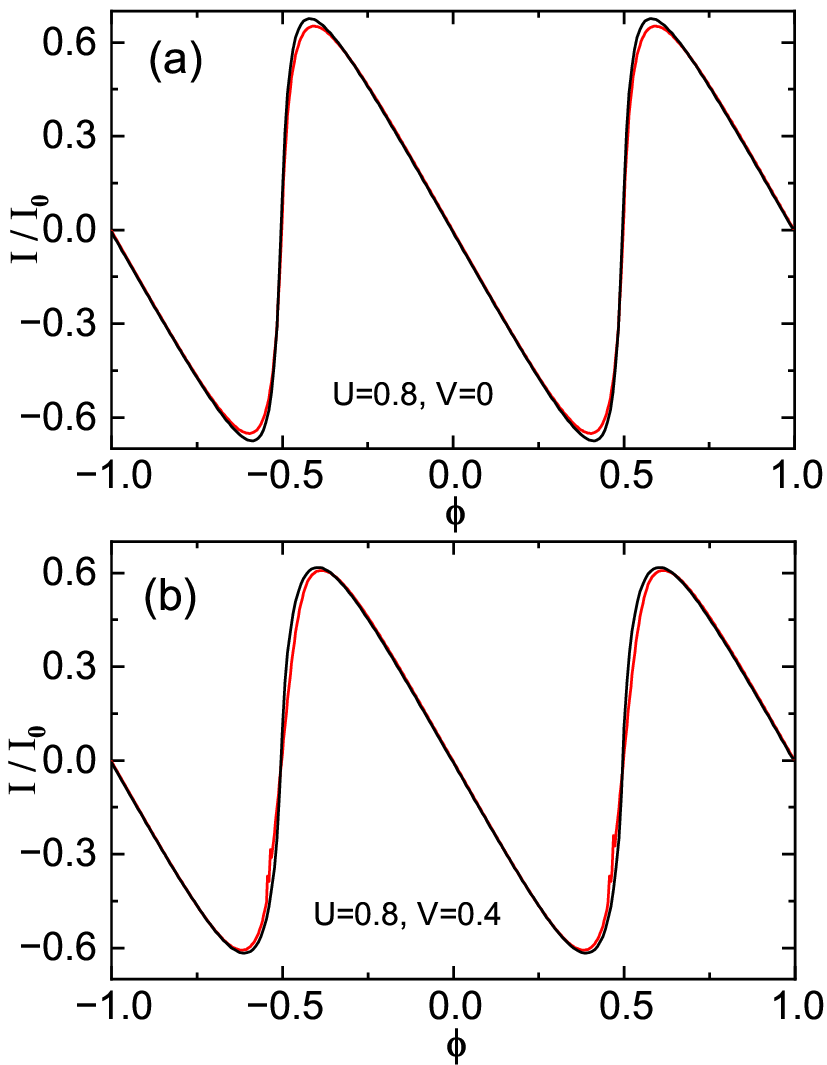}}\par}
\caption{(Color online). Comparison of persistent currents obtained from ED and MF analysis for a $10$-site half-filled
disordered ring with $W=1$. Panels $(a)$ and $(b)$ correspond to the interaction parameters $(U,V) = (0.8,0)$ and $(0.8,0.4)$, 
respectively. The black line represents the ED results, while the red line denotes the MF results. The flux $\phi$ is measured in units
of $\phi_0$.}
\label{comparison}  
\end{figure}
the corresponding current is evaluated following the expression given in Eq.~\ref{eqn6}. 

\vskip 0.2cm
\noindent
$\blacksquare$ {\bf Accuracy Check -- Comparison between ED and MF Results}: 
Before proceeding further, it is crucial to inspect the consistency and accuracy of the mean-field results in the presence of interaction as well as disorder. As illustrative examples, in Figs.~\ref{comparison}(a) and (b), we present a comparison of the persistent current obtained from ED and MF analysis for a $10$-site half-filled disordered ring ($W=1$), corresponding to interaction parameters $(U,V) = (0.8,0)$ and $(0.8,0.4)$, respectively, over the full flux range. In both cases, one with only on-site Hubbard interaction and the other with both on-site and nearest-neighbor Hubbard interactions, the resulting current profiles obtained from ED and MF methods almost perfectly overlap throughout the entire flux window. This excellent agreement clearly demonstrates that the MF scheme provides a reliable approximation. Therefore, we 
can safely employ this approach for further calculations. However, the agreement between the MF and ED results persists 
only over a limited parameter range. In the present work, we restrict ourselves to the regime $U < t$ and $V \leq U/2$.

\subsubsection{Dependence of PC with flux: larger ring system}

Figure~\ref{N=40, W=1, disorder average} illustrates the variation of the persistent current as a function of the magnetic flux $\phi$ for a representative disordered ring of size $N=40$ with disorder strength $W=1$, considering two distinct filling factors and different parameter sets. In Fig.~\ref{N=40, W=1, disorder average}(a), the black, red, and blue curves correspond to $U=0$, $0.3$, and $0.6$, respectively, with $V=0$ at half-filling. Figure~\ref{N=40, W=1, disorder average}(b) shows the results for $V=0.075$ (green curve) and $V=0.15$ (magenta curve), keeping $U=0.3$ under the same filling condition. 
\begin{figure}[ht]
\includegraphics[width=0.48\textwidth]{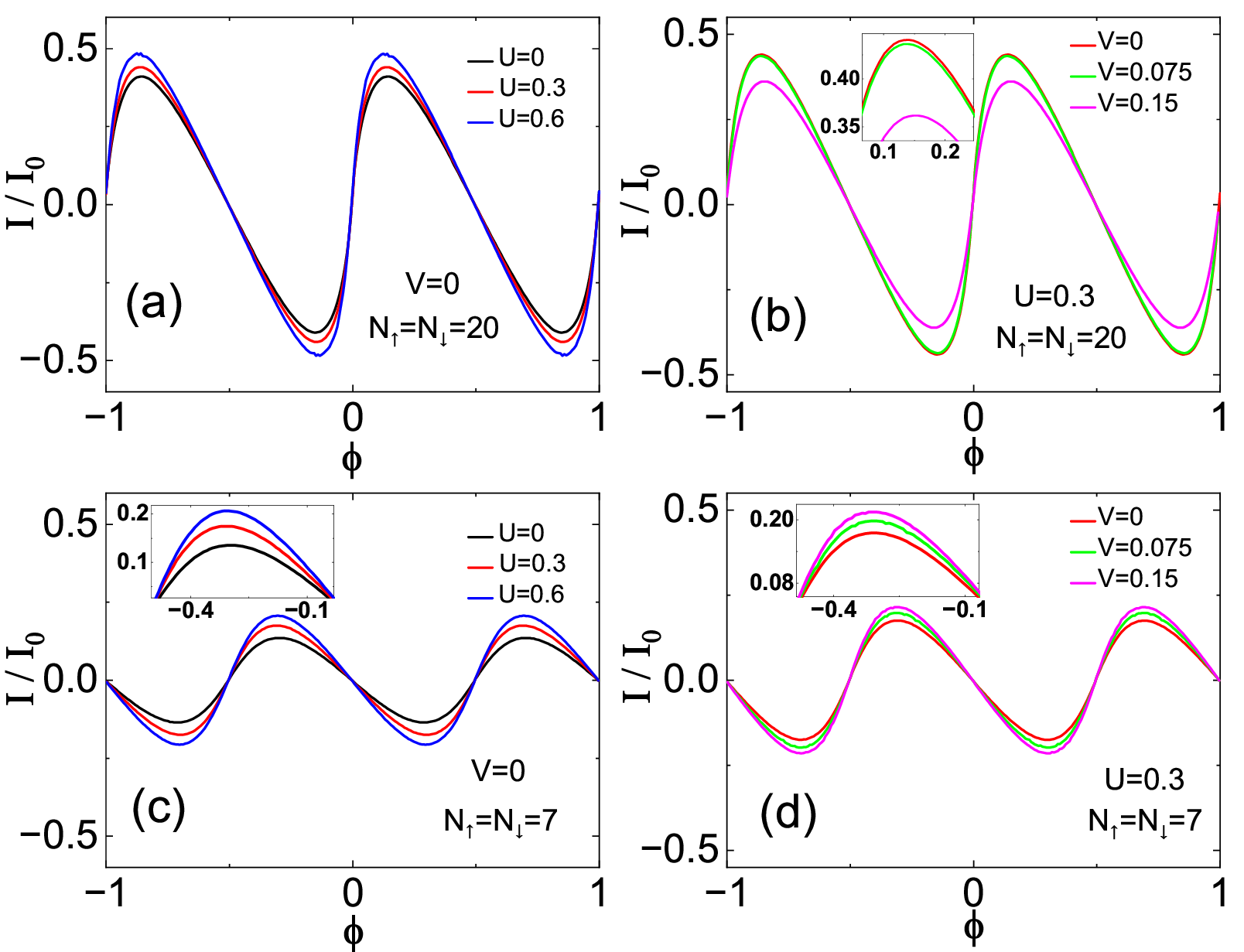}        
\caption{(Color online). Current-flux characteristics for a quantum ring of size $N=40$ with random disorder of strength $W=1$. The top row corresponds to the half-filled case ($N_\uparrow = N_\downarrow = 20$), while the bottom row represents a less-than-quarter-filled configuration ($N_\uparrow = N_\downarrow = 7$). All other relevant parameters for each curve are indicated within the figure. The flux is measured in units of $\phi_0$. Insets are included to highlight the current variations more clearly.}
\label{N=40, W=1, disorder average}     
\end{figure}
For a disordered ring at half-filling, we observe that the current increases with increasing $U$ but decreases upon introducing the interaction $V$.

The bottom row of Fig.~\ref{N=40, W=1, disorder average} presents the results for a less-than-quarter-filled case
($N_{\uparrow}=N_{\downarrow}=7$). In Fig.~\ref{N=40, W=1, disorder average}(c), the black, red, and blue curves represent the PC for $U=0$, $0.3$, and $0.6$, respectively, with $V=0$. In Fig.~\ref{N=40, W=1, disorder average}(d), the green curve shows the results for $V=0.075$ with $U=0.3$, while the magenta curve corresponds to $V=0.15$ for the same value of $U$. In this filling regime, the current increases with $U$, and it is further enhanced as $V$ is increased.

These results clearly demonstrate that, in the less-than-quarter-filled regime, increasing $V$ in the presence of $U$ enhances the persistent current, an opposite trend compared to higher fillings. Notably, this behavior fully persists even for larger system sizes and corroborates our findings obtained from the exact-diagonalization method.

\subsubsection{Dependence of PC with flux: different disorder strengths}

The results for disordered rings discussed so far have been obtained for a specific 
\begin{figure}[ht]
\includegraphics[width=0.48\textwidth]{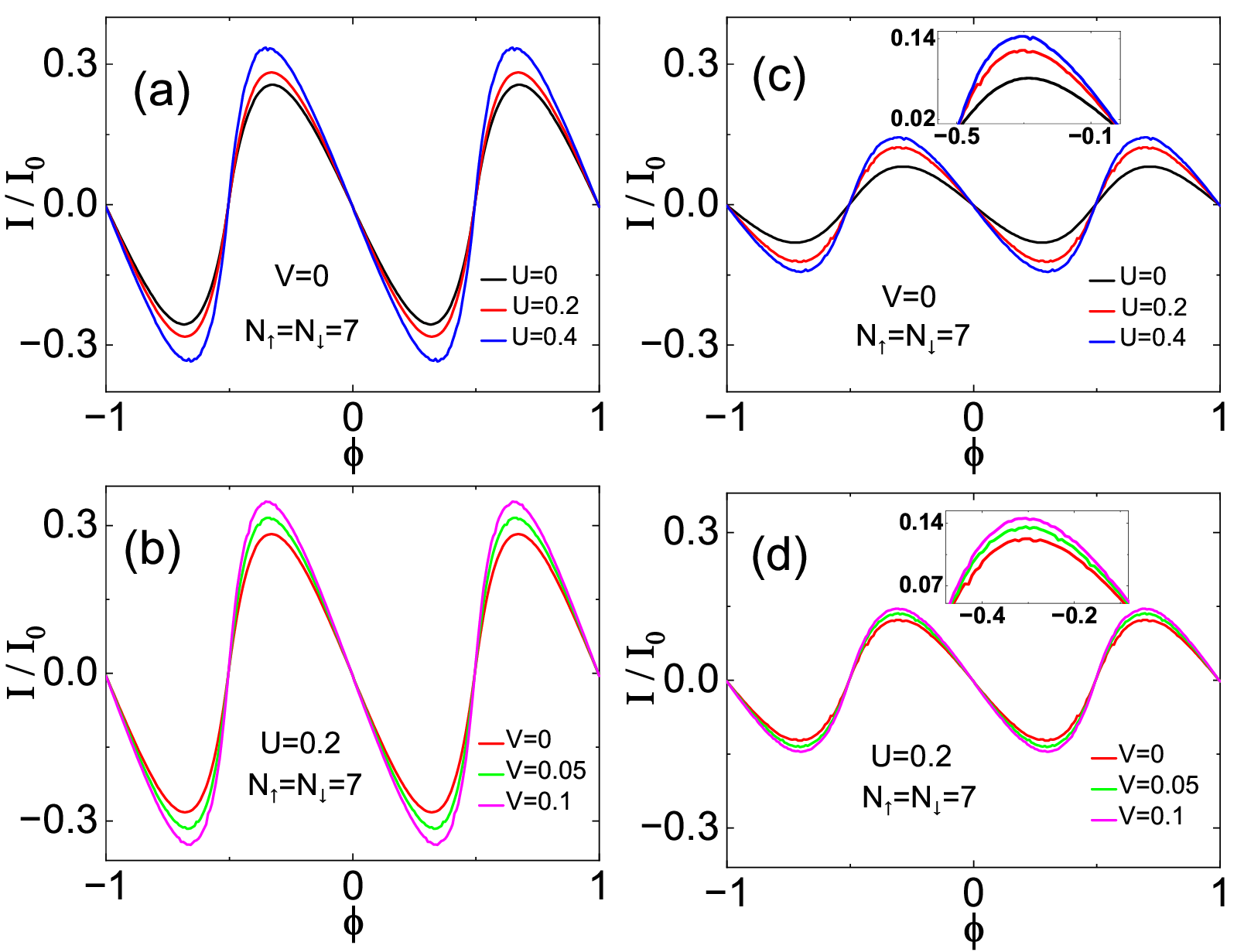}
\caption{(Color online). Current-flux characteristics for different disorder (random) strengths $W$ in a quantum ring of size $N=40$ with 
$N_\uparrow = N_\downarrow = 7$. The left and right columns correspond to $W=0.75$ and $W=1.25$, respectively. For the current variations 
with different values of $U$, the extended Hubbard strength $V$ is fixed to zero, whereas for the variations with different $V$ inputs, 
the on-site interaction is set to $U=0.2$. The flux is measured in units of $\phi_0$. Insets are provided to highlight the current variations 
more clearly.}
\label{fig:6}
\end{figure}
disorder strength, namely $W=1$. To verify the
consistency of our findings, we now examine two additional disorder strengths. The flux-driven persistent currents for $W=0.75$ and 
$W=1.25$ are presented in the left and right columns of Fig.~\ref{fig:6}, respectively, as representative examples. The ring size is 
kept identical to that used in Fig.~\ref{N=40, W=1, disorder average}, and random disorder is introduced in the site potentials.

Increasing the disorder strength enhances localization in the ring, following the well-known Anderson localization mechanism~\cite{al}.
Consequently, the persistent current diminishes with increasing disorder, in agreement with earlier studies~\cite{pc1,pc3}. As shown 
in Fig.~\ref{fig:6}, stronger disorder suppresses the peak magnitude of the current. For both disorder strengths, we observe that in 
the less-than-quarter-filled regime (here, $N_\uparrow = N_\downarrow = 7$ for a ring of size $N=40$), the amplitude of the persistent 
current increases in two cases: 
(i) when $U$ is increased while keeping $V$ fixed, and (ii) when $V$ is increased with $U$ held constant. This trend is fully consistent 
with the behavior reported earlier for $W=1$.

\subsubsection{Role of $U$ and $V$ on current magnitude from conducting behavior of eigenstates: An alternative analysis}

Here we present an alternative analysis to elucidate the crucial influence of the on-site and extended Hubbard interactions on the 
current magnitude. Our approach is based on examining how the spatial extent of individual energy eigenstates, quantified by the inverse
participation ratio (IPR), is modified by varying $U$ and $V$, since the conducting behavior of the system is inherently linked to the extendedness of its eigenstates, and thus the current.

\begin{figure}[ht]
	{\centering \resizebox*{8cm}{9cm}{\includegraphics{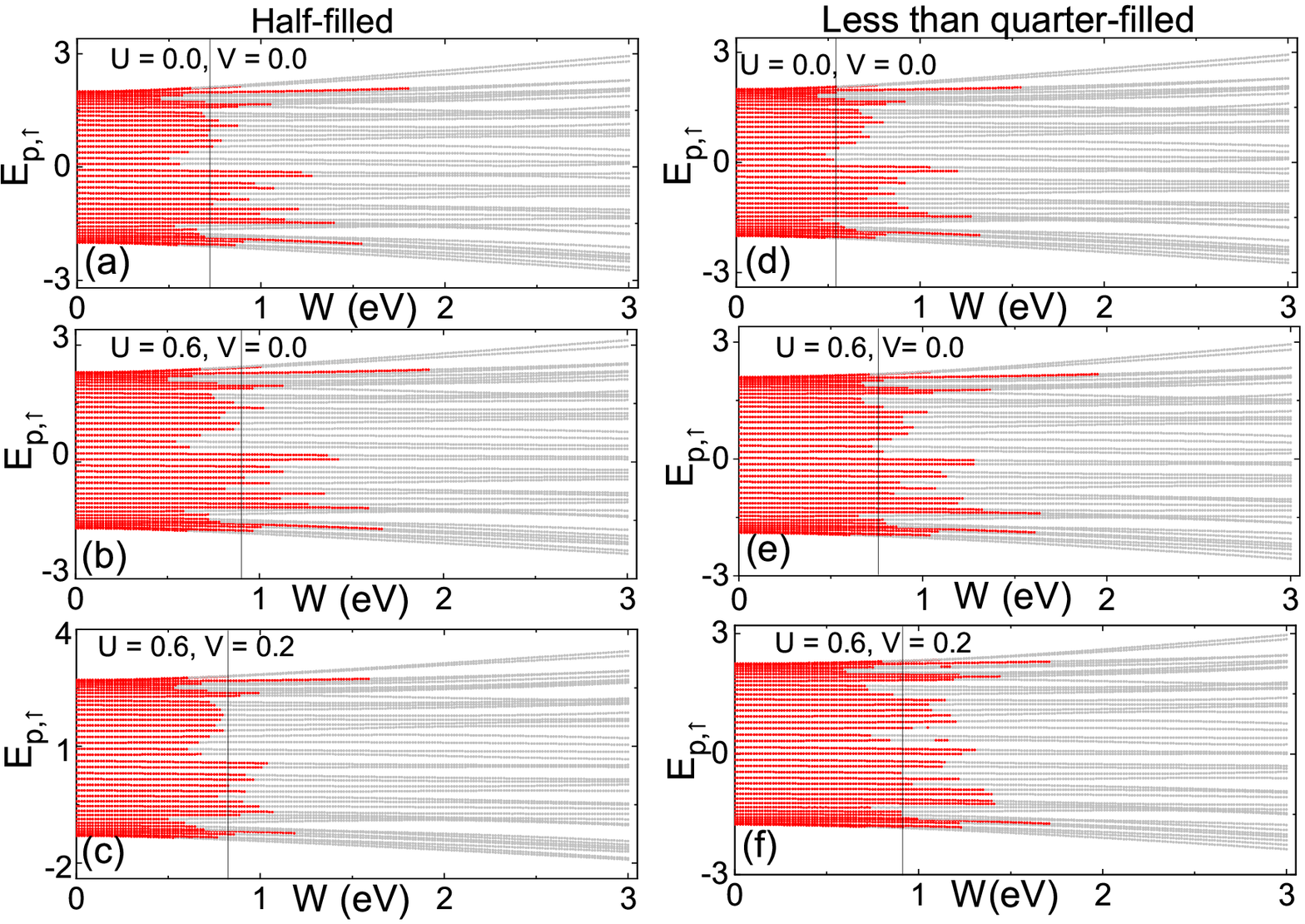}}\par}
	\caption{(Color online). Energy eigenvalue ($E_{p,\uparrow}=E_{p,\downarrow}$) spectrum of a $40$-site disordered (random) ring with 
		disorder strength $W$, when the flux is fixed at
		$0.75\phi_0$. Red dots indicate states with IPR $\leq 0.05$, while gray dots represent states with higher IPR values. The left column 
		displays the density profiles for various combinations of $U$ and $V$ in the half-filled regime, whereas the right column presents 
		the corresponding variations for a less-than-quarter-filled configuration with $N_{\uparrow} = N_{\downarrow} = 7$.}
	\label{fig:10}
\end{figure}

Figure~\ref{fig:10} shows the results for a $40$-site ring under different combinations of $U$ and $V$. The left column corresponds to 
the half-filled case, while the right column represents a less-than-quarter-filled scenario. Each sub-figure illustrates the variation of 
the IPR as functions of both the energy eigenvalues and the disorder realizations. Two colors are used to represent the IPR values: red 
for IPR $\le 0.05$, and gray for all higher values. 
In principle, an extended state is characterized by an IPR approaching zero, whereas 
a localized state exhibits an IPR approaching unity. Although these bounds cannot be exactly reached in a finite system, the relative 
magnitudes of IPR provide a reliable indication of extendedness. In our analysis, states with IPR $\le 0.05$ are considered extended, 
while those with larger values are treated as less conducting or effectively localized.

A vertical line is drawn in each sub-figure to demarcate the region up to which nearly all the states remain extended, beyond this 
point, the eigenstates rapidly become localized. In the half-filled case (Fig.~\ref{fig:10}(a-c)), this line shifts to the right upon 
increasing $U$ (Fig.~\ref{fig:10}(b)), indicating that stronger on-site repulsion enhances the extendedness of the eigenstates. This 
behavior is expected, as increasing $U$ effectively raises the average kinetic energy and promotes delocalization. However, when a
nearest-neighbor interaction is introduced ($U=0.6$, $V=0.2$), the line shifts leftward again (Fig.~\ref{fig:10}(c)), signaling 
enhanced localization. A stronger $V$ suppresses electron mobility, consistent with the reduction in current amplitude discussed earlier.

In contrast, the right column reveals a different trend for the less-than-quarter-filled regime. Here, the reference line moves to the 
right when $U$ is increased from $0$ to $0.6$ with $V=0$, showing enhanced extendedness, similar to the half-filled case. Interestingly,
unlike the half-filled situation, the line shifts even further to the right when the nearest-neighbor interaction is switched on 
($U=0.6$, $V=0.2$). The availability of many unoccupied sites in this low-filling regime allows electrons to hop more freely, thereby 
increasing their mobility. As a result, the eigenstates become more extended with increasing $V$, which directly explains the enhancement 
of current in this filling regime.

\section{Conclusion and Outlook}

In conclusion, we have systematically investigated the persistent current in ordered and disordered Hubbard rings under magnetic flux, 
employing both exact diagonalization and Hartree-Fock mean-field techniques within the tight-binding framework. Using a linear table 
formalism, we have constructed the full many-body Hamiltonian in an efficient and transparent manner, enabling ED calculations for 
system sizes that are otherwise challenging to access.

Our analysis reveals that the effects of on-site ($U$) and nearest-neighbor ($V$) interactions on the current are highly sensitive to 
electron filling and disorder. In ordered rings, the current decreases monotonically with increasing $U$, while the role of $V$ is 
nontrivial: it suppresses the current at low filling but enhances it near half-filling up to a critical ratio $V/U = 0.5$~\cite{order1}.
Disorder qualitatively modifies these behaviors, and we identify a distinct regime, less-than-quarter filling, where increasing $V$ leads 
to a significant enhancement of the current even in the presence of finite $U$. The inverse participation ratio (IPR) analysis of 
eigenstates further substantiates the interplay among $U$, $V$, filling factor, and disorder in controlling the current through 
modifications in the degree of localization.

Overall, our results establish clear parameter regimes in which extended Hubbard interaction can substantially enhance persistent current.
These regimes are directly accessible in experimentally realized extended Hubbard systems, such as quasi-one-dimensional 
materials like ET--F$_2$TCNQ~\cite{con_exp1,con_exp2}, where the extended interaction $V$ and hopping amplitude $t$ can be tuned and 
independently
extracted via pressure-dependent measurements. The formalism and analysis presented here can be extended to incorporate long-range hopping,
higher-order interactions, and finite-temperature effects~\cite{con1,con2}, providing a pathway toward a more comprehensive understanding 
of correlated electron transport in low-dimensional quantum systems. Beyond quantum electronic systems, such studies of correlation-induced
modifications of persistent current can also be relevant for ultracold atom platforms~\cite{atomtronics1,atomtronics2} within the framework 
of atomtronics~\cite{atomtronics3,atomtronics4}. These systems offer controllable environments to probe phase coherence and transport, 
thereby playing a crucial role in the design and optimization of devices under realistic interaction and disorder conditions.

\section*{ACKNOWLEDGMENT}

RS is grateful to CSIR, India (File number: 09/0093(17803)/2024-EMR-I) for providing his research fellowship.

\end{document}